\documentclass[twocolumn,prb,showpacs]{revtex4}
\usepackage{psfrag}
\usepackage{graphicx}
\usepackage{amsmath}
\usepackage{amssymb}

\begin{document}
\title{Fermi level alignment in molecular nanojunctions and its relation to charge transfer}
\author{R. Stadler}
\affiliation{Center for Atomic-scale Materials Physics, Department of Physics \\
  NanoDTU, Technical University of Denmark, DK-2800 Kgs. Lyngby, Denmark}
\author{K. W. Jacobsen}
\affiliation{Center for Atomic-scale Materials Physics, Department of Physics \\
  NanoDTU, Technical University of Denmark, DK-2800 Kgs. Lyngby, Denmark}

\date{\today}

\begin{abstract}
The alignment of the Fermi level of a metal electrode within the gap of the highest occupied (HOMO) and lowest unoccupied orbital (LUMO) of a molecule is a key quantity in molecular electronics, which can vary the electron transparency of a single molecule junction by orders of magnitude.
We present a quantitative analysis of the relation between this level alignment (which can be estimated from charging free molecules) and charge transfer for bipyridine and biphenyl dithiolate (BPDT) molecules attached to gold leads based on density functional theory calculations. For both systems the charge distribution is defined by a balance between Pauli repulsion with subsequent electrostatic screening and the filling of the LUMO, where bipyridine loses electrons to the leads and BPDT gains electrons. As a direct consequence the Fermi level of the metal is found close to the LUMO for bipyridine and close to the HOMO for BPDT.
 \end{abstract}
\maketitle

Interest for electron transport in nano-scale contacts has recently intensified, because (i) the advent of the technologically motivated field of molecular electronics~\cite{aviram74,eigler91,ohnishi98,joachim00} (ii) recent progress in the experimental techniques for manipulating and contacting individual molecules~\cite{joachim95,reed97,reichert02,smit02}, and (iii) the availability of first principles methods to describe the electrical properties of single molecule junctions.~\cite{transiesta,ratner,smeagol,calzolari04,fujimoto03} These latter methods are usually based on density functional theory (DFT) in combination with a non-equilibrium Green's function formalism\cite{keldysh65}.

\begin{figure*}
    \includegraphics[width=0.65\linewidth,angle=0]{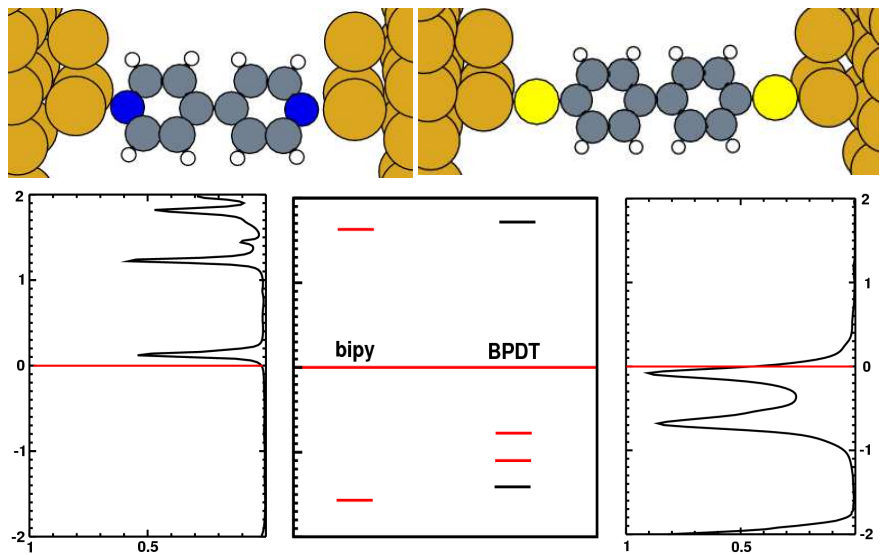}
      \caption[cap.Pt6]{\label{fig.intro} (Color online) A bipyridine (left) and BPDT (right) molecule suspended between Au electrodes. The middle lower panel shows the alignment of the molecular levels with the metals Fermi energy by equalizing vacuum potentials of the isolated molecules and surfaces where $E_F$, and the HOMO and LUMO are marked with gray (red) lines. The left and right lower panels show transmission functions for the coupled systems for bipyridine and BPDT, respectively.}
    \end{figure*}

It has become clear through many studies\cite{book} that the conductances of molecular junctions can be entirely controlled by the positions of individual molecular levels. In for example the case of a molecular contact consisting of a bipyridine molecule attached to Au leads it has been demonstrated\cite{xu,stadler}  that the transmission is completely dominated by the position of the lowest unoccupied molecular orbital (LUMO) and that this position may vary significantly with for example the surface structure of the leads. In order to describe the transport in molecular junctions it thus becomes a key issue to determine factors controlling the line-up of molecular levels relative to the Fermi level of the metal. Is the alignment governed mostly by the decay of the potential at the surface or it is controlled by level shifts due to the specific interaction between a particular molecule and a particular surface? In order to approach this question we study in the following two different molecular junctions which turn out to behave quite differently with respect to level alignment.

Fig.~\ref{fig.intro} compares bipyridine and biphenyl dithiolate (BPDT) attached to gold electrodes with the same surface structure and the same bonding configuration (top panels). In both cases no molecular levels can be found close to the Fermi energy if the vacuum potentials from separate calculations for the molecules and the metal surface are set equal (middle lower panel). However, when the molecules are coupled to the electrodes, there are peaks in the transmission functions (left and right lower panels), where $E_F$ is crossed by their tails at different sides of a gap for the two different molecules. 
For bypridine, the LUMO has moved downwards with respect to its original (non-interacting) position, for BPDT it has moved upwards. The exact position of these peaks determines the conductance which is defined by the value of the transmission function at $E_F$ and can vary between 0.03 and 0.44 $G_0$ for BPDT depending on the bonding configuration. 
What we shall show in the following is that the level shifts can be directly determined from an appropriately defined charge transfer between the metal surface and the molecule. The down-shift of the bipyridine LUMO is for example associated with an electronic charge transfer from the molecule to the surface. This might at first seem counterintuitive since the down-shift of the LUMO apparently leads to a slight occupation of this state (see Fig.~\ref{fig.intro}), but still the net electron transfer is away from the molecule due to a combination of Pauli repulsion and screening effects as we shall demonstrate. For benzene dithiolate (BDT) and BPDT the alignment of molecular levels coupled to a Au (111) surface has previously been suggested to be linked to charge being transferred from the surfaces to the molecules.~\cite{xue1}
 
All electronic structure calculations in this study are performed using a plane wave implementation of DFT~\cite{dacapo} with an energy cutoff of 340 eV, where we used ultra-soft pseudopotentials~\cite{vanderbilt90}, and a PW91 parametrization for the exchange and correlation functional.~\cite{pw91} 
The transmission functions of the molecular junctions in Fig.~\ref{fig.intro} were calculated using a general non-equilibrium Green's function formalism for phase-coherent electron transport~\cite{meir92}, where both, the Green's function of the scattering region and the self-energies describing the coupling to the semi-infinite electrodes, were evaluated in terms of a basis consisting of maximally localized Wannier functions obtained from a transformation of the Kohn-Sham eigenstates~\cite{thygesen} within an energy range up to 2 eV above E$_F$. In our calculations the supercells for the scattering region are defined by $3\times 3$ atoms in the directions perpendicular to the transport direction and contain three to four surface layers on each side of the molecule. We found that a $4\times 4$ $\bold k$-point grid is needed for the sampling in the transverse Brillouin plane in order to obtain well converged results for the conductance~\cite{kpoints}.

  \begin{figure}   
  \includegraphics[width=0.95\linewidth,angle=0]{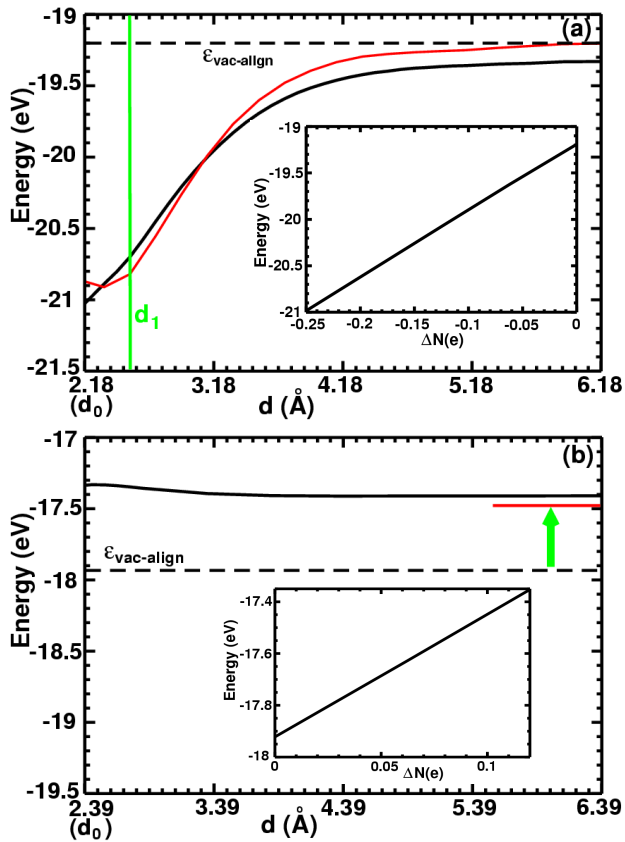}   
  \caption[cap.Pt6]{\label{fig.bipy_trans}(Color online) MO1 energies (relative to E$_F$ of the Au surface) depending on the distance $d$ between the molecule and the Au surfaces for a) bipyridine and b) BPDT. The black solid curves are taken from the coupled systems, for the red curves estimates from the charged free molecules have been used, and the dashed line shows the position from vacuum alignment without charging ($\epsilon_{vac-align}$). MO1 in its dependence on the charge for the free molecules is shown as insets. 
}    
  \end{figure}     

We investigated the variation of the energetic position of the lowest-lying molecular orbital (denoted MO1 in the following text) with respect to the metal's Fermi level, in dependence on the distance between the surface and the molecule $d$. This is depicted in Fig.~\ref{fig.bipy_trans}, where $d_{0}$ marks the equilibrium bond length between the nitrogen atoms of the bipyridine molecule (Fig.~\ref{fig.bipy_trans}a) or the sulfur atoms of BPDT (Fig.~\ref{fig.bipy_trans}b) and the Au atoms they are attached to.
Since MO1 is $\sim$ 10 eV below the lowest-lying Au valence states, its energetic position must be exclusively guided by rigid potential shifts without any direct hybridisation effects. 
Now we want to address the question whether these rigid potential shifts have a quantifiable relation to charge being transferred between the molecule and the surface.
In our study we make use of the concept of fractional charges, which can be introduced either through a straight forward extension of the mathematical framework of DFT or via a thermodynamic ensemble interpretation of the mixing of pure states.~\cite{casida} This makes it possible to determine the ground state electron density and electronic eigenvalue spectrum for a (albeit only finite) system with fractions of electrons removed or added when compared to the total charge of all the nucleis. The shift of MO1 in the charged isolated molecules is shown as insets in Fig.~\ref{fig.bipy_trans} for bipyridine and BPDT. As can be seen the molecular levels move up when the molecule is charged as a consequence of electrostatic repulsion. Our main argument is that these level shifts of the free molecule can explain the level shifts in the coupled molecule surface system. To show this we use the charge density differences between the coupled and isolated subsystems and the linear energy/charge relation for the free molecule. This allows us to calculate the position of MO1 as a function of distance, which we then compare directly with the actual energies from the coupled system in Fig.~\ref{fig.bipy_trans}. 
For a physical interpretation of such fractional charges two notes of caution have to be made: i) Fractional charges are only physical meaningful for small distances $d$, where an equilibrium between the molecule and the surface can be assumed. Our plots for long $d$ are only used for simulating a gradual {\it switching on} of the interaction. ii) In our calculations we do not correct for the lack of the derivative discontinuity (DD) of the exchange correlation potential within standard DFT methods, which can lead to an underestimation of the HOMO-LUMO gap and an overestimation in the conductance for weakly coupled single molecule junctions even at low biases~\cite{burke}. Including DD would make our scheme impractical, since MO1 would jump discontinously when going from negative to positive charges. However, our study does not focus on absolute values for conductance or gap size but on the comparison of equilibrium charge transfer in two different junctions.

For bipyridine at large distances $d$, MO1 rests at an energetic position $\epsilon_{vac-align}$, which corresponds to the one it would hold if the vacuum levels of the isolated molecule and surface were aligned (Fig.~\ref{fig.bipy_trans}). This indicates that there is no interaction between the two subsystems and that both are entirely charge neutral for $d$ larger than $\sim$ 6~\AA\ . At the bonding distance $d_0$ an effective charge of -0.25 electrons on the bipyridine can be derived from the shift of MO1 by comparison with the charged free molecule. Partial charges with respect to the isolated subsystems can also be computed directly from electron density differences, which results in a charge of -0.23 electrons on the bipyridine molecule at the same distance.
As can be seen from Fig.~\ref{fig.intro}, the HOMO-LUMO gap is much smaller for BPDT ($\sim$ 0.3 eV) than for bipyridine ($\sim$ 3.2 eV). This is because BPDT lacks two electrons, which are subtracted from its $\pi$ system, when two hydrogen atoms are removed from the stable aromatic molecule biphenyl dithiol in order to form the highly reactive biradical BPDT, which is then attached to the Au surfaces. Since the molecular levels corresponding to the dangling bonds on the sulfur atoms are fully occupied in BPDT, the HOMO of biphenyl dithiol gets emptied and becomes the LUMO of BPDT.
A further difference between bipyridine and BPDT (see Fig.~\ref{fig.bipy_trans}) is that the long distance position of MO1 for BPDT is not just $\epsilon_{vac-align}$ but is instead $\sim$ 0.5 eV higher in energy. Vacuum level alignment for BPDT on Au (111) leads to a situation where the LUMO lies well below $E_F$ (see Fig.~\ref{fig.intro}), and independent of $d$ the two subsystems cannot exist within the same cell without charge being transferred from the surface to the molecule.
We find 0.11 electrons gain on the BPDT molecule from the MO1 level shifts and 0.09 from charge density differences $\Delta n(x_z)$ at large $d$. At the bonding distance $d_0$, the BPDT molecule gains 0.12 electrons calculated from MO1 level shifts and $\Delta n(x_z)$ cannot be interpreted unambigously due to the strong hybridisation of Au and S states. We note that BPDT differs from bipyridine in that charge transfer occurs in the opposite direction, where the amount of charge being moved is by almost a factor of two smaller for BPDT and its dependence on $d$ is negligible.

\begin{figure}      
\includegraphics[width=0.95\linewidth,angle=0]{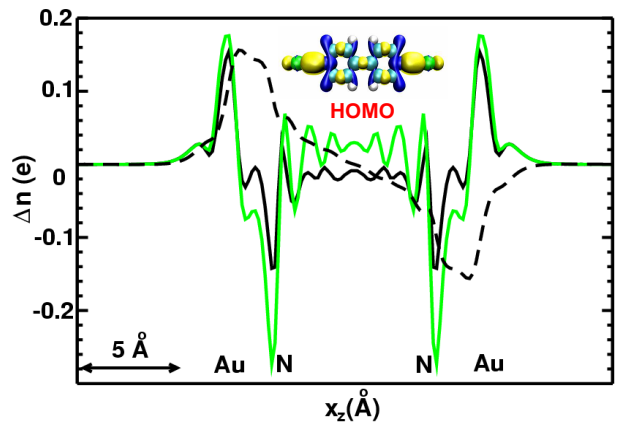}      
\caption[cap.Pt6]{\label{fig.bipy_comp} (Color online) Charge density difference $\Delta n(x_z)$ (black solid line, summed up parallel to the surface plane) and its integral (black dashed line) for bipyridine coupled to Au atoms ($d$=2.63~\AA\ ). $\Delta n(x_z)$ constructed only from contributions of the bipyridine HOMO (see inset) and the Au s and d ($z^2$) states is also given for comparison (green line).
}    
\end{figure} 

An apparent problem with the picture we propose here is the following: In the case of bipyridine charge is moving away from the molecule and the levels are therefore shifting down in energy. However, as the LUMO hits E$_F$ it must begin to fill and this is naturally associated with charge being transferred to the molecule. Both effects are in fact taking place. For the further analysis of the bipyridine junction, we use a model system where the leads are replaced by single Au atoms. 
Looking at $\Delta n(x_z)$ in Fig.~\ref{fig.bipy_comp} we find maxima at the Au atoms and minima close to but not at the nitrogen positions. When we form charge density differences just using the bipyridine HOMO (which is twice degenerate and both orbitals are fully occupied), the Au s and the also fully occupied d ($z^2$) states, the result still has the same nodal structure as $\Delta n(x_z)$. The two functions differ in the minimum at the nitrogens being deeper for the difference constructed only from six orbitals. In the latter case there is also no charge depletion in the center of the molecule. Both aspects can be explained in terms of screening. Since a very localised minimum in the charge density is energetically unfavourable, the lower lying MOs are polarised so that this minimum is at least partially smoothed out. Such a polarisation of MOs has the effect that the total charge density at the center of the molecule is reduced, which explains the net charge transfer from the molecule to the Au. For the maximum at the Au atoms no such screening occurs since for the bare atoms there are no electrons available for achieving that. For the realistic surface calculations, however, the situation is different and the same peak in $\Delta n(x_z)$ (not shown here) becomes smaller and broader. 

But what is the effect of the partial filling of the LUMO, which occurs in the bipyridine junction only for small $d$? It can be seen in the change of curvature in the MO1 energies vs. $d$ in Fig.~\ref{fig.bipy_trans}a for distances smaller than $d_1$. Because now the process described in the last paragraph is partially neutralised by charge flowing back to the molecule, the lowering in energy of MO1 is slightly reduced, thereby flattening the curves. For Fig.~\ref{fig.bipy_comp} we have chosen a distance larger than $d_1$ in order to be able to discuss the two driving forces for charge movement separately. 
For BPDT the same balance between Pauli repulsion of electrons (leading to electron depletion on the molecule) and the filling of the LUMO exists but the latter process dominates, and the net effect is therefore a movement of electrons towards the molecule. The strong system dependence of the Fermi level alignment in molecular junctions can also be seen by a comparison of our results with a recent article on alkane thiolates on Au (111)~\cite{scandolo}, where the dominant effect was found to be due to molecular dipoles which result from the asymmetry of the system.

In summary, we presented a detailed analysis of the energetic alignment of molecular orbitals with respect to the Fermi level of gold electrodes in single molecule nanojunctions with bipyridine and biphenyl dithiolate molecules. The outcome of this alignment has a crucial effect on the zero bias conductance of the junction, the major source determining it is equilibrium charge transfer between the molecule and the electrodes. 
We established that a comparison between the energies of the lowest lying molecular orbitals within the junction and for the isolated but partially charged molecule gives reasonable estimates for the net charge being transferred. For the bipyridine, the molecule is drained of electrons, the biphenyl dithiolate on the other hand gains electrons from the Au electrodes. 
From a simplified model of the junctions where the electrodes have been replaced by single Au atoms, we could derive that the charge transfer process for the two molecules we studied is determined by the balance of two effects. Pauli repulsion between occupied molecular and Au orbitals pushes electrons away from the molecule and the interface region to more remote parts of the electrodes surface or bulk, which followed by screening depletes the molecules of electronic charge. The filling of the LUMO on the other hand results in an electron surplus on the molecule. For bipyridine the first effect dominates, for BPDT the situation is reversed.

We appreciate useful discussions with Kristian Thygesen and Mikkel Strange. The Center for Atomic-scale Materials Physics at NanoDTU is sponsored by the Danish National Research Foundation. We acknowledge support from the Nano-Science Center at the University of Copenhagen and from the Danish Center for Scientific Computing through Grant No. HDW-1101-05.


\bibliographystyle{apsrev}

\begin{thebibliography}{23}
\expandafter\ifx\csname natexlab\endcsname\relax\def\natexlab#1{#1}\fi
\expandafter\ifx\csname bibnamefont\endcsname\relax
  \def\bibnamefont#1{#1}\fi
\expandafter\ifx\csname bibfnamefont\endcsname\relax
  \def\bibfnamefont#1{#1}\fi
\expandafter\ifx\csname citenamefont\endcsname\relax
  \def\citenamefont#1{#1}\fi
\expandafter\ifx\csname url\endcsname\relax
  \def\url#1{\texttt{#1}}\fi
\expandafter\ifx\csname urlprefix\endcsname\relax\def\urlprefix{URL }\fi
\providecommand{\bibinfo}[2]{#2}
\providecommand{\eprint}[2][]{\url{#2}}

\bibitem{ohnishi98}
  H.~Ohnishi, Y.~Kondo, and K.~Takayanagi
  Nature (London) {\bf 344}, 524 (1998).
\bibitem{joachim00}
  C.~Joachim, J.~K.~Gimzewski and A.~Aviram
  Nature (London) {\bf 408}, 541 (2000).
\bibitem{aviram74}
  A.~Aviram and M.~A.~Ratner,
  Chem. Phys. Lett. {\bf 29}, 277 (1974).
\bibitem{eigler91}
  D.~M.~Eigler and E.~K.~Schweizer
  Nature {\bf 344}, 56 (1991).

\bibitem{joachim95}C.~Joachim, J.~K.~Gimzewski, R.~R.~Schlittler and
  C.~Chavy Phys. Rev. Lett. {\bf 74}, 2102 (1995).

\bibitem{reed97}M.~A. Reed,
  C.~Zhou, C.~J. Muller, T.~P. Burgin and J.~M. Tour Science {\bf
    278}, 252 (1997).

\bibitem{reichert02}J.~Reichert, R. Ochs, D. Beckman, H.~B.  Weber,
  M. Mayor and H.~v. L{\"ohneysen} Phys. Rev. Lett. {\bf 88}, 176804
  (2002).

\bibitem{smit02}R.~H.~M.~Smit, Y.~Noat, C.~Untiedt, N.~D.~Lang, M.~C.~van
  Hemert and J.~M.~van Ruitenbeek, Nature {\bf 419}, 906 (2002).

\bibitem{fujimoto03}
  Y.~Fujimoto and K.~Hirose
  Phys. Rev. B {\bf 67}, 195315 (2003).
\bibitem{smeagol}
  A.~R.~Rocha, V.~M.~Garc\'{i}a-Su\'{a}rez, S.~W.~Baily, C.~J.~Lambert, J.~Ferrer and S.~Sanvito
  Nature Materials {\bf 4}, 335 (2005).
\bibitem{transiesta}
  M.~Brandbyge, J.~L.~Mozos, P.~Ordej\'{o}n, J.~Taylor, and K.~Stokbro
  Phys. Rev. B {\bf 65},165401 (2002).
\bibitem{ratner}
  Y.~Xue, S.~Datta, and M.~A.~Ratner
  Chem. Phys {\bf 281}, 151 (2002).
\bibitem{calzolari04}
  A.~Calzolari, N.~Marzari, I.~Souza, M.~B.~Nardelli
  Phys. Rev. B {\bf 69}, 35108 (2004).
\bibitem{keldysh65}
  L.~V.~Keldysh
  Soviet Physics JETP-USSR {\bf 20}, 1018 (1965).
\bibitem{book}G.~Cuniberti, G.~Fagas, and K. Richter (Eds.), {\it Introducing Molecular Electronics} (Springer, 2005).
\bibitem{stadler} R.~Stadler, K.~S.~Thygesen, and K.~W.~Jacobsen, Phys. Rev. B {\bf 72}, 241401(R) (2005).
\bibitem{xu} B.~Xu and N.~J.~Tao, Science {\bf 301}, 1221 (2003); B.~Xu, X.~Xiao and N.~J.~Tao, J. Am. Chem. Soc. {\bf 125}, 16164 (2003).

\bibitem{xue1}Y.~Xue, S.~Datta and M.~A.~Ratner, J.~Chem.~Phys. {\bf 115}, 4292 (2001); Y.~Xue and M.~A.~Ratner, Phys.~Rev.~B {\bf 68}, 115406 (2003).

\bibitem{dacapo} 
  B. Hammer, L.B. Hansen, and J.K. N{\o}rskov,
  Phys.\ Rev.\ B {\bf 59}, 7413 (1999); 
  S.R. Bahn and K.W. Jacobsen, 
  Comp.\ Sci.\ Eng.\ {\bf 4}, 56 (2002); 
  The Dacapo code can be downloaded at http://www.fysik.dtu.dk/campos.
\bibitem{vanderbilt90}
  D.~Vanderbilt, 
  Phys. Rev. B {\bf 41}, R7892 (1990).
\bibitem{pw91}
  J.~P. Perdew {\it et~al.}, Phys. Rev. B {\bf 46}, 6671
  (1992). 
\bibitem{meir92}
  Y.~Meir and N.~S.~Wingreen,
  Phys. Rev. Lett. {\bf 68}, 2512 (1992).
\bibitem{thygesen}K.~S.~Thygesen and K~ W.~Jacobsen, Chem. Phys. {\bf 319}, 111 (2005); K.~S.~Thygesen, L.~B.~Hansen and K.~W.~Jacobsen, Phys. Rev. Lett. {\bf 94}, 026405 (2005); K.~S.~Thygesen, L.~B.~Hansen and K.~W.~Jacobsen, Phys. Rev. B {\bf 72}, 125119 (2005).
\bibitem{kpoints} K.~S.~Thygesen and K~ W.~Jacobsen, Phys. Rev. B {\bf 72}, 0334 01 (2005).
\bibitem{casida}M.~E.~Casida, Phys. Rev. B {\bf 59}, 4694 (1999).
\bibitem{burke} C.~Toher, A.~Filippetti, S.~Sanvito and K.~Burke, Phys. Rev. Lett. {\bf 95}, 146402 (2005); M.~Koentopp, K.~Burke and F.~Evers, Phys. Rev. B {\bf 73}, 121403 (2006).
\bibitem{scandolo}R.~Rousseau, V.~De~Renzi, R.~Mazzarello, D.~Marchetto, R.~Biagi, S.~Scandolo and U.~del~Pennino, J. Phys. Chem. B, in print (2006).
\end{thebibliography}

\end{document}